\documentstyle[prb,aps,floats,epsfig]{revtex}

\begin{document}
\twocolumn[\hsize\textwidth\columnwidth\hsize\csname @twocolumnfalse\endcsname

\title{Violation of particle number conservation in the $GW$ approximation}
\author{Arno Schindlmayr\cite{email}}
\address{Cavendish Laboratory, University of Cambridge, Madingley Road,
Cambridge CB3 0HE, United Kingdom}
\date{Received 5 February 1997}
\maketitle

\begin{abstract}
We present a nontrivial model system of interacting electrons that can be
solved analytically in the $GW$ approximation. We obtain the particle
number from the $GW$ Green's function strictly analytically, and prove that
there is a genuine violation of particle number conservation if the
self-energy is calculated non-self-consistently from a zeroth order Green's
function, as done in virtually all practical implementations. We also show
that a simple shift of the self-energy that partially restores
self-consistency reduces the numerical deviation significantly.
\end{abstract}

]
\narrowtext

\section*{Introduction}

Many-body perturbation theory for condensed-matter physics allows the
Green's function of a system of interacting electrons to be formulated in
a picture of independent quasiparticles moving in an effective potential.
The key quantity that incorporates the contributions of dynamic exchange
and correlation to this effective potential is the self-energy operator
$\Sigma$, which in general is both nonlocal and energy dependent. It is
itself a functional of the Green's function that can formally be
expressed through an infinite series of Feynman diagrams. In practice,
however, approximations for the functional form are required, of which the
most popular is the $GW$ approximation introduced by Hedin.\cite{Hed65}
This approach replaces the infinite series of Feynman diagrams for the
self-energy operator by a single Fock-like diagram that is the product of
the Green's function $G$ and the dynamically screened Coulomb interaction
$W$, hence the name of the scheme. Originally the $GW$ approximation was
devised to be applied self-consistently on an equal footing with the
Hartree and Hartree-Fock approximations, in the sense that the Green's
function used to generate the self-energy be identical to that obtained
from it, but the computational cost proved prohibitive. Thus $GW$
self-energies are traditionally calculated from a zeroth order Hartree or
LDA Green's function, although a small degree of self-consistency is
sometimes introduced by rigidly shifting the zeroth order Green's function
on the energy axis in such a way that its chemical potential agrees with
that of the $GW$ Green's function derived from it. It is in this fashion
that the $GW$ approximation has in the past been successfully applied to
a wide range of materials including semiconductors,\cite{Hyb85,God86}
simple metals,\cite{Nor87} and transition metals.\cite{Ary92} The first
fully self-consistent $GW$ calculations for model systems were reported only
recently,\cite{Gro95,Shi96} but these consistently showed an undesired loss
of structural features in the Green's function in disagreement with
experiment.\cite{Bar96}

As future $GW$ calculations for realistic materials are therefore unlikely
to adopt a fully self-consistent approach, the question of particle number
conservation, long since a subject of debate in this context, has gained
renewed significance. Since the imaginary part of the Green's function
is directly related to the density of states as well as to the real-space
charge density, it must integrate to the correct number of electrons
contained in the system, a condition frequently used to determine the
numerical accuracy in practical implementations. However, while the fully
self-consistent $GW$ scheme is known to conserve the exact particle
number,\cite{Bay61} the same, while tacitly assumed, has never actually been
proven for the approximation used in practice, in which the self-energy is
calculated from a zeroth order Green's function. Numerical simulations
appeared to corroborate this positive assumption, with deviations below one
percent fully within the range expected due to systematic numerical errors,
and this has already prompted occasional confirmative conjectures in the
literature.\cite{Ary92} However, in this paper we shall show that upon
elimination of these errors, which arise from numerical integration,
transformation to Fourier space with a finite broadening of the
quasiparticle peaks, alignment of the chemical potentials only within
second order perturbation theory, etc., there still remains a genuine
albeit numerically small violation of particle number conservation, as was
previously demonstrated by accurate molecular calculations for the
analogous scheme based on time-dependent Hartree-Fock rather than Hartree
theory that is frequently employed in quantum chemistry.\cite{Res76} We
also demonstrate that restoring a degree of self-consistency by means of an
appropriate shift in the zeroth order Green's function, which in practice
is often omitted because of its negligible influence on the band structure,
further significantly reduces the deviation. The significance of this result
is that it provides a way to minimize the fundamental limit on the
accuracy with which quantities such as the charge density or the total
energy can be calculated in the $GW$ approximation.

\section*{Description of model system}

For our proof we consider as an analytically solvable counter-example a
four-site Hubbard cluster with tetrahedral symmetry, populated by two
electrons. To the best of our knowledge this is in fact the first model
system forwarded in the literature for which the Green's function and
particle number in the $GW$ approximation can be calculated strictly
analytically, without any numerical errors. Its significance for theoretical
investigations therefore stretches beyond the objective of this paper, and
for the benefit of the reader we will thus present all derivations in
sufficient detail. The Hamiltonian is
\begin{equation}
{\cal H} = \left(\! \epsilon - \frac{U}{2} \right) \sum_{{\bf R},\sigma}
\hat{n}_{{\bf R}\sigma} -t \!\!\sum_{{\bf R},{\bf R}',\sigma}\!\!
c^\dagger_{{\bf R}\sigma} c_{{\bf R}'\sigma} + U \sum_{\bf R}
\hat{n}_{{\bf R}\uparrow} \hat{n}_{{\bf R}\downarrow} ,
\end{equation}
where $c^\dagger_{{\bf R}\sigma}$ and $c_{{\bf R}\sigma}$ are the creation
and annihilation operators for an electron at site ${\bf R}$ with spin
$\sigma$, and $\hat{n}_{{\bf R}\sigma} = c^\dagger_{{\bf R}\sigma}
c_{{\bf R}\sigma}$. We set $\epsilon = 4 t$ and $t = 1$ while leaving the
onsite interaction strength $U$ variable. To construct the $GW$ self-energy
we start from a zeroth order Green's function in the Hartree approximation,
noting that the exchange-correlation potential in a corresponding
density-functional treatment is a mere constant due to spatial symmetries
and would not affect the following arguments. Spatial symmetries and
degeneracy in $\sigma$ also require a uniform fractional site occupation of
one half and accordingly a Hartree potential of $U/2$ on all sites.
Analytic diagonalization of the Hartree $4 \times 4$ Hamiltonian matrix
${\cal H}^{\rm H}_{{\bf RR}'} = \epsilon \delta_{{\bf RR}'} - t$ for each
spin orientation then yields a nondegenerate ground state at zero energy and
a threefold degenerate excited state at energy $\epsilon$. With the electrons
in the ground state $| 0 \rangle$ the zeroth order Green's function becomes
\begin{equation}
G^{\rm H}_{{\bf RR}'}(\omega) = \frac{\langle {\bf R} | 0 \rangle \langle 0
| {\bf R}' \rangle}{\omega - i\eta} + \sum_{\nu=1}^3 \frac{\langle {\bf R} |
\nu \rangle \langle \nu | {\bf R}' \rangle}{\omega - \epsilon + i\eta} ,
\end{equation}
where we have omitted the spin index. $\eta$ denotes a positive
infinitesimal. Using the relation $\langle {\bf R} | 0 \rangle \langle 0 |
{\bf R}' \rangle = 1/4$ for the components of the ground-state vector we
proceed to calculating the polarization propagator in the random-phase
approximation (RPA), defined through
\begin{eqnarray}
P^{\rm RPA}_{{\bf RR}'}(\omega)
&=& -2 \frac{i}{2\pi} \int\!\! G^{\rm H}_{{\bf RR}'}(\omega+\omega')
G^{\rm H}_{{\bf R}'{\bf R}}(\omega) \,d\omega' \nonumber \\
&=& \frac{1}{2} \sum_{\nu=1}^3 \langle {\bf R} | \nu \rangle
\langle \nu | {\bf R}' \rangle \!\left\{ \frac{1}{\omega - \epsilon + i\eta}
- \frac{1}{\omega + \epsilon - i\eta} \right\} , \nonumber \\
&& \mbox{}
\end{eqnarray}
including a factor 2 for the spin summation. The polarization propagator is
diagonal in eigenvector space, and can thus be analytically inverted to
yield the screened interaction
\begin{eqnarray}
W^{\rm RPA}_{{\bf RR}'}(\omega)
&=& U \delta_{{\bf RR}'} + U \sum_{{\bf R}''} P^{\rm RPA}_{{\bf RR}''}(\omega)
W^{\rm RPA}_{{\bf R}''{\bf R}'}(\omega) \nonumber \\
&=& U \delta_{{\bf RR}'} + \frac{\epsilon U^2}{2 z} \sum_{\nu=1}^3
\langle {\bf R} | \nu \rangle \langle \nu | {\bf R}' \rangle \nonumber \\
&& \times \left\{ \frac{1}{\omega - z + i\eta}
- \frac{1}{\omega + z - i\eta} \right\} ,
\end{eqnarray}
with $z = [\epsilon (\epsilon + U)]^{1/2}$. A similar analytic expression
for the screening in a related model system was previously given in Ref.\
\onlinecite{Ver95}. The self-energy in the $GW$ approximation is a convolution
of the zeroth order Green's function and the RPA screened interaction,
\begin{eqnarray} \label{sigma}
\Sigma^{GW}_{{\bf RR}'}(\omega)
&=& \frac{i}{2\pi} \int\!\! G^{\rm H}_{{\bf RR}'}(\omega-\omega')
W^{\rm RPA}_{{\bf RR}'}(\omega') e^{i \delta \omega'} \,d\omega' \nonumber \\
&=& -\frac{U}{4} \delta_{{\bf RR}'} + \frac{3 \epsilon U^2}{8 z}
\frac{\langle {\bf R} | 0 \rangle \langle 0 | {\bf R}' \rangle}{\omega - z
- \epsilon + i\eta} \nonumber \\
&&+ \frac{\epsilon U^2}{8 z} \sum_{\nu=1}^3 \langle {\bf R} | \nu \rangle
\langle \nu | {\bf R}' \rangle \nonumber \\
&& \times \left\{ \frac{2}{\omega - z - \epsilon + i\eta}
+ \frac{1}{\omega + z - i\eta} \right\} ,
\end{eqnarray}
where $\delta$ denotes a positive infinitesimal. We have written
$\Sigma^{GW}$ in such a way as to emphasize that it is diagonal in the
eigenvectors $| \nu \rangle$ of the initial Hartree system. On the other
hand, diagonalization of the full Hamiltonian including the self-energy
yields the quasiparticle states of the interacting electron system, so for
our model the two are in fact identical, although the corresponding energy
eigenvalues are not. In this way we can calculate the chemical potential
$\mu$ of the interacting electron system exactly from its true quasiparticle
properties. By definition the chemical potential is identical to the energy
eigenvalue of the highest occupied quasiparticle state, which is zero for
the original Hartree system, and at the $GW$ level is given implicitly
through the self-energy correction,
\begin{equation}
\mu = 0 + \Sigma^{GW}_{\nu=0}(\mu-\tilde{\omega}) .
\end{equation}
Here we have allowed for the possibility of using a self-energy derived
from a zeroth order Green's function whose chemical potential has been
shifted by $\tilde{\omega}$ on the energy axis. To simulate the effect of
a self-consistent calculation we determine the shift by requiring the
chemical potentials of the shifted zeroth order Green's function and the
$GW$ Green's function obtained from it to be identical. While this equation
is usually solved within second order perturbation theory, the simple form
of the self-energy for our model (\ref{sigma}) allows us to derive the
exact analytic solution
\begin{equation} \label{shift}
\tilde{\omega} = -\frac{U}{4} - \frac{3 \epsilon U^2}{8 z (z+\epsilon)} ,
\end{equation}
which correctly approaches zero as $U \to 0$. Accordingly we solve Dyson's
equation
\begin{equation} \label{dyson}
G^{GW}_{\nu}(\omega) = G^{\rm H}_{\nu}(\omega) + G^{\rm H}_{\nu}(\omega)
\Sigma^{GW}_{\nu}(\omega-\tilde{\omega}) G^{GW}_{\nu}(\omega)
\end{equation}
in eigenvector space both for $\tilde{\omega} = 0$ and $\tilde{\omega}$ as
in Eq.\ (\ref{shift}), and compare the results. In this diagonal form
Dyson's equation is analytically solvable. For $\nu = 0$ the self-energy
(\ref{sigma}) contains one pole, which adds a satellite to the
quasiparticle peak of the Hartree Green's function and yields a quadratic
equation for the positions of the poles of $G^{GW}$. Similarly the
quasiparticle and satellite structure of the $\nu > 0$ matrix elements is
obtained from the zeroes of a third order polynomial, reflecting the richer
spectrum in the self-energy. The $GW$ Green's function therefore takes the
analytic form
\begin{eqnarray} \label{greensfun}
\lefteqn{ G^{GW}_{{\bf RR}'}(\omega+\tilde{\omega}) } \nonumber \\
&=& \langle {\bf R} | 0 \rangle \langle 0 | {\bf R}' \rangle
\frac{\omega-z-\epsilon}{(\omega-x_1-i\eta) (\omega-x_2+i\eta)} \nonumber \\
&&+ \sum_{\nu=1}^3 \langle {\bf R} | \nu \rangle \langle \nu | {\bf R}'
\rangle \nonumber \\
&&\times \frac{(\omega-z-\epsilon) (\omega+z)}{(\omega-y_1-i\eta)
(\omega-y_2+i\eta) (\omega-y_3+i\eta)} ,
\end{eqnarray}
where we have defined the symbols
\begin{equation}
x_{1,2} = \frac{\displaystyle z+\epsilon-\tilde{\omega}-\frac{U}{4}}{2}
\mp \sqrt{ \left( \frac{\displaystyle z+\epsilon+\tilde{\omega}
+\frac{U}{4}}{2} \right)^2 + \frac{3 \epsilon U^2}{8z} }
\end{equation}
together with $y_1 = -(b/3) - 2r \cos(\phi/3)$ and $y_{2,3} = -(b/3) + 2r
\cos[(\pi \mp \phi)/3]$ as well as the auxiliary quantities
$r = \pm \sqrt{|p|}$, $\phi = \arccos(q/r^3)$, $q = b^3/27 - bc/6 + d/2$,
$p = (3c-b^2)/9$, and the polynomial coefficients
\begin{mathletters}
\begin{eqnarray}
b &=& -2\epsilon + \tilde{\omega} + \frac{U}{4} , \\
c &=& -z (z+\epsilon) + \epsilon \left( \epsilon-\tilde{\omega}
-\frac{U}{4} \right) - \frac{3 \epsilon U^2}{8z} , \\
d &=& z (z+\epsilon) \left( \epsilon-\tilde{\omega}-\frac{U}{4} \right)
+\frac{(\epsilon - z) \epsilon U^2}{8z} .
\end{eqnarray}
\end{mathletters}

\section*{Results}

\begin{figure}
\epsfxsize=3.25in \epsfbox{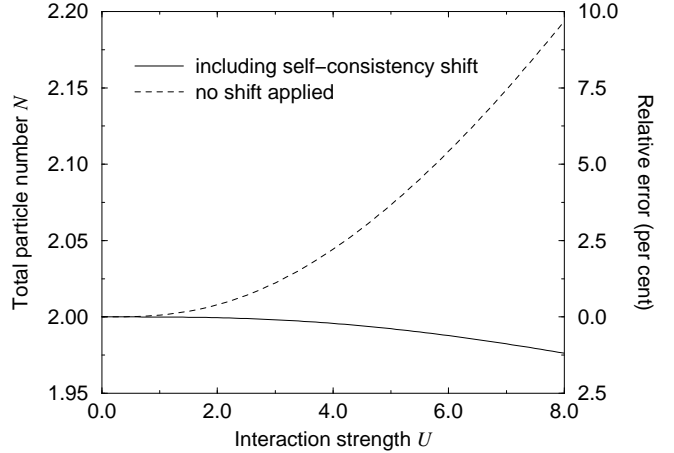}
\caption{Analytically calculated particle number in the $GW$ approximation
for our two-electron model system as a function of the interaction strength
$U$. Applying a rigid shift to the zeroth order Green's function such as to
align its chemical potential self-consistently with that of the $GW$
Green's function significantly reduces the violation of particle number
conservation, but fails to remove it completely.}
\label{fig:particlenumber}
\end{figure}

From the Green's function in the $GW$ approximation (\ref{greensfun}) the
particle number may be obtained by an analytic contour integration along a
path closed across the upper half plane, to sample all occupied states
below the chemical potential. By inspection we note that these are the
states at $x_1$ and $y_1$. For the total particle number we thus obtain
\begin{eqnarray} \label{number}
N &=& 2 \sum_{\bf R} \frac{1}{2 \pi i} \int\!\! G^{GW}_{\bf RR}(\omega)
e^{i \delta \omega} \,d\omega \nonumber \\
&=& \frac{2 (x_1-z-\epsilon)}{x_1-x_2} + \frac{6 (y_1-z-\epsilon)
(y_1+z)}{(y_1-y_2) (y_1-y_3)} ,
\end{eqnarray}
including a factor 2 for spin summation. If the particle number in the $GW$
approximation was conserved, $N$ would have to be a constant with a value
of two. In particular, it would also have to be independent of the
interaction strength $U$, which so far we have not specified, but an
analysis of the expression (\ref{number}) confirms that this is not the
case. In Fig.\ \ref{fig:particlenumber} we show the calculated particle
number as a function of $U$, both with and without applying the shift
$\tilde{\omega}$ in the zeroth order Green's function.
In either case the growing deviation from the true value as $U$ increases
is clearly visible, evidence of a genuine and fundamental violation
of particle number conservation in the $GW$ approximation as usually
applied. In absolute terms, however, we find that the discrepancy is much
reduced when the chemical potentials are aligned in the prescribed way, and
that it depends only very weakly on the interaction strength up to high
values of $U$. For a medium correlation of $U = 4$ the numerical deviation
amounts to an underestimation of merely 0.21\% of the true particle number.

A comparison of the weight factors of $G^{GW}$ with those of the exact
Green's function, calculated by numerical diagonalization of the Hamiltonian
matrix, shows that in the absence of the self-consistency shift
$\tilde{\omega}$ the main error in $N$ stems from a serious overestimation of
the satellite spectrum, while the weight of the quasiparticle peak at $x_1$
deviates relatively little from the correct value up to high correlation
strength. When the self-consistency shift is applied the satellites are
also overestimated, although by a lesser amount. In this case, however, this
is the result of a balanced weight transfer from the quasiparticle peaks that
has little influence on the integrated spectral weight and yields a total
particle number in much better agreement with the correct value.

\section*{Summary}

In summary, we have presented a two-electron model system for which the
particle number from the Green's function in the $GW$ approximation can be
derived strictly analytically, without any additional inaccuracies that have
previously beset numerical calculations. Through an analysis of this model
system we have demonstrated that there is a genuine violation of particle
number conservation in the $GW$ approximation as it is usually applied,
i.e., with the self-energy calculated non-self-consistently from a zeroth
order Green's function. However, we also find that the numerical deviation
from the exact particle number can be kept low even for an extremely strong
correlation by introducing a small degree of self-consistency in the form
of a simple rigid shift of the zeroth order Green's function on the energy
axis in such a way as to align its chemical potential with that of the $GW$
Green's function, which is calculated from it through Dyson's equation.
While the effect of this shift on the quasiparticle band structure may be
small, our results clearly indicate its significance for the calculation of
quantities such as the particle number and charge density that are derived
through integration over the complete spectral function.

\acknowledgements

The author wishes to thank R.~W.\ Godby, M.~M.\ Rieger, and T.~J.\ Pollehn
for useful discussions, and acknowledges financial support from the
Deut\-scher Aka\-de\-mi\-scher Aus\-tausch\-dienst under its HSP III scheme,
the Stu\-di\-en\-stif\-tung des deut\-schen Vol\-kes, the Gott\-lieb
Daimler- und Karl Benz-Stiftung, Pembroke College Cambridge, and the
Engineering and Physical Sciences Research Council.

\end{document}